\documentclass[aps, prd, preprint, superscriptaddress, tightenlines, nofootinbib]{revtex4}

\usepackage{amssymb, amsmath} 
\usepackage{stmaryrd}

\usepackage{bbm} 

\usepackage[english]{babel}
\usepackage{slashed}
\usepackage{cancel}

\usepackage{slashbox}

\usepackage{graphicx}
\usepackage{color}

\usepackage{datetime}
\settimeformat{ampmtime}

\usepackage{float}

\usepackage[bookmarks,breaklinks,plainpages=false,pdfpagelabels]{hyperref} 
	\hypersetup{linkcolor=red,citecolor=blue,filecolor=dullmagenta,urlcolor=darkblue} 

\usepackage{soul}

\definecolor{grey}{rgb}{0.4,0.4,0.4}
\definecolor{lightgrey}{rgb}{0.6,0.6,0.6}
\definecolor{dullmagenta}{rgb}{0.4,0,0.4}
\definecolor{darkblue}{rgb}{0,0,0.4}
\definecolor{orange}{rgb}{1,0.5,0}
\definecolor{lightbrown}{rgb}{0.75,0.5,0.25}
\definecolor{tan}{cmyk}{0.14,0.42,0.56,0}
\definecolor{djunglegreen}{cmyk}{0.99,0,0.52,0}
\definecolor{lightgreen}{rgb}{0,1,0}
\definecolor{olivegreen}{cmyk}{0.64,0,0.95,0.40}
\definecolor{midgreen}{rgb}{0.0,0.675,0.0}
\definecolor{darkgreen}{rgb}{0,0.5,0}

\newcommand{\vs}{\vspace}

\renewcommand{\.}{\hspace{0.5mm}}

\newcommand{\irm}{\ensuremath{\mathrm{i}}}

\renewcommand{\d}{\ensuremath{\mathrm{d}}}
\newcommand{\ee}{\ensuremath{\mathrm{e}}}

\newcommand{\defas}{\mathrel{\mathop :}=} 
\newcommand{\hph}[1]{\hphantom{#1\;\,}}

\newcommand{\eg}{e.g.}
\newcommand{\ie}{i.e.}

\newcommand{\cf}{cf.}

\makeatletter
 \def\ifundefined#1{\expandafter\ifx\csname#1\endcsname\relax}
 \ifundefined{default@color}
  \let\default@color=\current@color
 \fi
\makeatother

\newcommand{\beq}{\begin{equation}}
\newcommand{\eeq}{\end{equation}}
\newcommand{\bea}{\begin{eqnarray}}
\newcommand{\beas}{\begin{eqnarray*}}

\newcommand{\eea}{\end{eqnarray}}
\newcommand{\eeas}{\end{eqnarray*}}

\begin{document}

\title{High-Energy Gravitational Scattering and Bose-Einstein Condensates of Gravitons}

\author{Florian K{\"u}hnel}
\email{florian.kuhnel@fysik.su.se}
\affiliation{The Oskar Klein Centre for Cosmoparticle Physics,
	Department of Physics,
	Stockholm University,
	AlbaNova,
	106\.91 Stockholm,
	Sweden}

\author{Bo Sundborg}
\email{bo.sundborg@fysik.su.se}
\affiliation{The Oskar Klein Centre for Cosmoparticle Physics,
	Department of Physics,
	Stockholm University,
	AlbaNova,
	106\.91 Stockholm,
	Sweden}

\date{\formatdate{\day}{\month}{\year}, \currenttime}

\begin{abstract}
Quantum black holes are difficult to describe. We consider two seemingly divergent approaches, high-energy scattering and the proposal to regard black holes as Bose-Einstein condensates of gravitons, and establish a connection between them. High-energy scattering is studied in the eikonal approximation, which is processed further by a saddle-point approximation. The dominant contribution to the scattering amplitude comes from a ladder diagram with the exchange of $N$ gravitons, and the number of gravitons follows a Poisson distribution. This approximation supports the picture of a graviton Bose-Einstein condensate with an extent equal the Schwarzschild radius, which grows with $N$ in a way determined by the saddle point. The approach permits calculations of $1 / N$ corrections from the fluctuations around the saddle points and we comment on these. Scattering methods might be useful probes of quantum black holes, especially when interpreted in terms of condensates.
\end{abstract}


\maketitle

\section{Introduction}
\label{sec:Introduction}

\vs{-2mm}
The quantum mechanics of black holes is one of the big puzzles in physics. Many conceptual problems, like the information paradox and the no-hair theorem, become apparent when a classical black-hole solution is combined with quantum field theory in such a background. Studies of such approximations are clearly relevant to our understanding of the classical limit of black holes, but there are less explored paths, which might lead more directly to a description of quantum black holes. Recently, Dvali and Gomez \cite{Dvali} have proposed a picture of black holes as a Bose-Einstein condensate of a large number, $N$, of gravitons. A number of essential black-hole features have been matched by estimates in this language such as Bekenstein entropy and Hawking temperature (\cf~also \cite{N-Portrait-related, Brustein, Casadio}). Such a picture can potentially improve our understanding conceptually, but also quantitatively, if $1 / N$ corrections to the leading behaviour can be evaluated.

High-energy scattering in theories containing gravity provides another potential route to black-hole physics. The idea is essentially that the increase of the effective gravitational coupling with energy can lead to strong fields and the formation of black holes, for instance if two scattering particles approach one another sufficiently closely on the scale of the Schwarzschild radius corresponding to the total energy of the process. More concretely, string theory raised the hope that loop diagrams in gravitational theories could make sense and perhaps that the loop expansion of the high-energy scattering can be resummed using an eikonal approximation, the faster growth with energy of higher-order diagrams notwithstanding \cite{Amati, Sundborg:1988tb, Muzinich:1987in}. Almost immediately it was realized that the dominant contributions at small angle and high energy are independent of short distance and ultraviolet (UV) completions like string theory, and this line of ideas has later been explored further (\cf~\cite{Giddings}, for a pedagogical review see \cite{Giddings:2011xs}).\footnote{There have also been developments on how to extract non-universal effects dependent on the string UV completion \cite{D'Appollonio:2010ae}.}

A possible reaction to these observations is to elevate them to the level of a principle: Theories like gravity may protect themselves from UV problems by channelling energy into large numbers of low energy quanta \cite{t'Hoof1987}. The corresponding classical configurations has recently been called studied much further under the name of classicalons \cite{Dvali:2010jz, Dvali:2011th} (\cf~also earlier works related to so-called ``self-completion of gravity'' \cite{Dvali:2010bf}).

The eikonal resummation of the high-energy gravitational scattering gives an uncontroversial but puzzling result (see below). A simple physical picture is wanting, and any such picture is in need of quantitative support.

In previous work \cite{Kuhnel:2014zja} we have observed a nice and clean relation between higher-dimensional black holes, Bose-Einstein condensates and classicalons. This leads us to believe that such relations will also show up directly in high-energy scattering, and lead to a clear interpretation of the eikonal amplitude. The present paper checks this idea and takes some step further by investigating the fluctuations around the saddle points that dominate the physics. It is our hope that such fluctuations can play an important role in the study of quantum corrections to the leading classical description of large black holes.
Much of the physics of the eikonal approximation, of classicalons as black holes and graviton condensates is well known, and the main role of this work is to focus on and spell out the strength of the relationships between these ideas in as simple a setting as possible. We demonstrate aspects of how they fruitfully support each other, and find new insights in the effective distribution of exchanged gravitons.

This work is structured as follows: In Sec.~\ref{sec:Eikonal-Approximation} we introduce the eikonal approximation, in a way directly applicable to the studied set-up. We then apply a saddle-point approximation to obtain the energy scaling of the impact parameter. In Sec.~\ref{sec:Around the Characteristic Impact Parameter} we analyse fluctuation around this saddle point. Sec.~\ref{sec:The-Coherent-Sum-over-Gravitons} concerns with the determination of the dominant contribution amongst the ladder-diagrams in the eikonal sum, where we also determine the distribution of exchanged gravitons at the saddle point. We conclude and summarise in Sec.~\ref{sec:Summary-and-Discussion}.

\section{Eikonal Approximation}
\label{sec:Eikonal-Approximation}

\begin{figure}
	\centering
	\includegraphics[scale=0.52,angle=90]{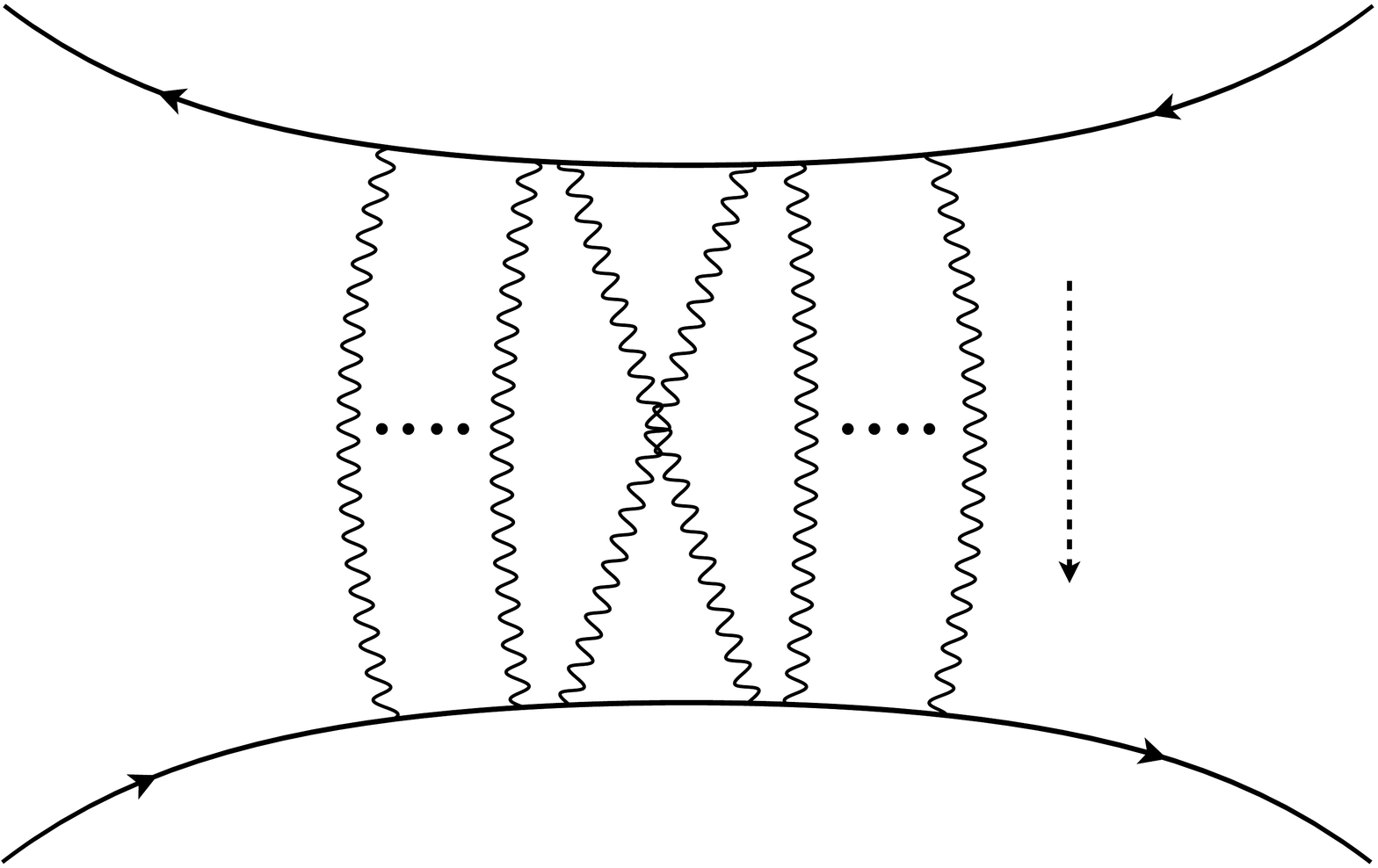}
	\caption{A typical ladder diagram which contributes to the sum \eqref{eq:exponent}.}
	\label{fig:Ladder-Diagram}
	\begin{picture}(0,0)
		\put(125,195){$q$}
	\end{picture}
\end{figure}

We consider high-energy scattering in $D$ space-time dimensions. Let $b$ be the impact parameter of the scattering process, $q \defas p_{1} - p_{2}$ the momentum transfer, and $s$ the square of the center-of-mass energy of the gravitons with initial $D$-momenta $p_{1}$ and $p_{2}$. Let us further assume small momentum transfer,
\begin{align}
	q
		&\ll
								\sqrt{s\.}
								\; .
								\label{eq:q<<sqrt(s)}
\end{align}
Now, the eikonal amplitude $A^{\text{E}}$, which can be expressed as
\begin{align}
	A^{\text{E}}( s, q )
		&\sim
								s
								\int\!\d^{ D - 2 } b\; \ee^{\irm q\cdot b}\;
								\frac{ \ee^{\irm \chi( s, b )} - 1 }{ 2 \irm }
								\; ,
								\label{eq:Eikonal-Amplitide}
\end{align}
can be regarded as an estimate of the sum of ladders and crossed ladders in the $s$-channel (\cf~\cite{Collins}). In Fig.~\ref{fig:Ladder-Diagram} we depict a typical relevant diagram. The eikonal function $\chi(s, b)$ can be expressed in terms of the Born amplitude\footnote{Here taken from string theory to explicitly show how string contributions are subdominant.} $A^{(0)}$,
\begin{align}
	A^{(0)}( s, q )
		&\sim
								-
								c\.\kappa_{D}^{2}\.
								\frac{ s^{2} }{ q^{2} }\.
								\ee^{ -a q^{2} }
								\; ,
								\label{eq:Born-Amplitude}
\end{align}
via
\begin{subequations}
\begin{align}
	\chi(s, b)
		&\sim
								\frac{ 2 }{ s ( 2 \pi )^{D - 2} }
								\int\!\d^{ D - 2 } q\;
								\ee^{-\irm q \cdot b}\.
								A^{(0)}( s, q )
								\\[4mm]
		&\sim
								-
								\frac{ 4\.s\.c\.\kappa_{D}^{2}\. }{ \sqrt{4\.\pi\.}^{D - 2}\.\sqrt{a\.}^{D - 4}\.(D - 4) }\;
								{}^{}_{1}F^{}_{1}
								\bigg(
									\frac{ 1 }{ 2 }\.( D - 4 );\,
									\frac{ 1 }{ 2 }\.( D - 2 );\,
									-\frac{ b^{2} }{ 4 a }
								\bigg)
								\; .
								\label{eq:Eikonal-Function}
\end{align}
\end{subequations}
Above, $c$ is some real constant, $\kappa_{D}$ is the $D$-dimensional gravitational constant, ${}^{}_{1}F^{}_{1}$ is the confluent hyper-geometric function of the first kind, and we defined $a \defas \ln( s ) - \ln( 2 ) - \irm\.\pi / 2$. Now, for $b \gg a$ we find
\vs{-1mm}
\begin{align}
	\chi(s, b)
		&\sim
								-
								\frac{ s\.c\.\kappa_{D}^{2} }{ \sqrt{\pi\.}^{D - 2}\.(D - 4) }\,
								\Bigg[
									\frac{ \Gamma
										\big[
											( D - 2 ) / 2
										\big] }
									{ b^{D - 4} }
									-
									\frac{ ( D - 4 )\.\ee^{- b^{2} / 4 a} }
									{ 2\.\sqrt{4\.a\.}^{D - 6}\,b^{2} }\.
								\Bigg]
								\, ,
								\label{eq:Eikonal-Function-for-b>>a}
\end{align}
where the second term in the delimiters represents string contributions \cite{Sundborg:1988tb}. For large $b$ this term can safely be neglected, and we find for the eikonal amplitude
\begin{align}
\begin{split}
	A^{\text{E}}( s, q )
		&\sim
								-
								\frac{ s^{( D - 1 ) / 2 } }{ \sqrt{2\.\pi\.} }
								\int^{\infty}\!\d b\;
								b^{( D - 3 ) / 2}\.
								\cos\!
								\big[
									q\.b
									-
									\tfrac{ \pi }{ 4 }( D - 3 )
								\big]\.
								\times
								\\[1mm]								
		&\hph{\sim}
								\phantom{-\frac{ s^{( D - 1 ) / 2 } }{ \sqrt{2\.\pi\.} }\int\!\d b\;\;\;}
								\times
								\frac{ s\.c\.\kappa_{D}^{2}\.
									\Gamma
									\big[
										( D - 2 ) / 2
									\big] }
									{ \sqrt{\pi\.}^{D - 2}\.b^{D - 3} }\,
								\exp\!
								\Bigg[
									-
									\irm\.
									\frac{
									s\.c\.\kappa_{D}^{2}\.
									\Gamma
									\big[
										( D - 2 ) / 2
									\big] }
									{ ( D - 4 )\sqrt{\pi\.}^{D - 2}\.b^{D - 4} }
								\Bigg]
								\, .
\end{split}
								\label{eq:Born-Amplituide-large-b}
\end{align}
This oscillatory integral can be approximated by its value at the saddle point $b_{*}$. Assuming a small scattering angle $\theta \sim q / \sqrt{s\.}$, we have
\begin{align}
	b_{*}
		&\sim
								\bigg[
									\frac{s}{\theta^{2}}
								\bigg]^{\frac{ 1 }{ 2 D - 6 }}
		\sim						r_{\text{S}}\,
								\theta^{-\frac{ 1 }{D - 3}}
								\; ,
								\label{eq:bstar}
\end{align}
where $r_{\text{S}}$ is the Schwarzschild radius of a black hole with mass $M = \sqrt{s\.}$. This growth of the typical impact parameter with energy is thus the same as the growth of the Schwarzschild radius of a black hole (\cf~\cite{Kuhnel:2014zja}, including generalizations to classicalons). However the impact parameter is scaled with a large number depending on the small angle $\theta$ probing the system. It is clear that small angle probes of the formation of black holes in high-energy scattering safely avoid the actual horizon. If we insert the saddle-point value of $b$ in Eq.~\eqref{eq:Born-Amplituide-large-b} and evaluate $A^{\text{E}}( s, q )$, we get a decent estimate, but as a function of $s$ and $q$ (see, \eg, Ref.~\cite{Sundborg:1988tb}) it is far from transparent and in need of an interpretation.

\subsection{Around the Characteristic Impact Parameter}
\label{sec:Around the Characteristic Impact Parameter}

A necessary requirement for the saddle-point approximation to make sense is a large second derivative of the exponent in Eq.~\eqref{eq:Born-Amplituide-large-b} at the saddle point $b_{*}$,
\begin{align}
	\frac{ \partial^{2} \chi}{ \partial b^{2} } \Big|_{b\.=\.b_{*}}
		&\sim
								s\.b_{*}^{ D - 2 }
		\sim
								s^{ \frac{ D - 4 }{ 2D - 6 } }\.
								\theta^{ \frac{ D - 2 }{ D - 3 } }
								\; .
								\label{eq:second-derivative-of-the-exponent-at-the-saddle-point}
\end{align}
For all space-time dimensions $D > 4$ we find that the exponent of $s$ is non-negative, and is monotonically increasing with $D$. This implies that the saddle-point approximation is getting better and better at higher energies, and that the improvement with energy is faster in higher dimensions. The typical impact parameter is more sharply defined at higher energies, as seen by evaluating the width of the saddle-point Gaussian
\begin{align}
	\Delta b
		&\sim	
								s^{- \frac{ D - 4 }{ 4(D - 3) } }\.
								\theta^{- \frac{ D - 2 }{ 2(D - 3) } }
								\; .
								\label{eq:width-at-the-saddle-point}
\end{align}

Unfortunately, this does not work equally well in the $D = 4$ limit, but still the relative width $\Delta b / b_{*} \to 0$. That the $D = 4$ case is special is already apparent from Eq.~\eqref{eq:Eikonal-Function-for-b>>a}, where the limit $D \to 4$ gives a constant divergence plus a $\log b$ dependence. The divergence translates to phase by Eq.~\eqref{eq:Eikonal-Amplitide}, but the constant phase is unobservable and related to the very long-range nature of the $1 / r$ potential in $D = 4$. We regard $D = 4$ as a special case that requires special technical care but which is unlikely to be qualitatively different than $D > 4$.

\subsection{The Coherent Sum over Gravitons}
\label{sec:The-Coherent-Sum-over-Gravitons}

Let us go back to the eikonal amplitude $A^{\text{E}}$, given in Eq.~\eqref{eq:Born-Amplituide-large-b}. As mentioned earlier, the exponential in $A^{\text{E}}$ sums all the ladder-diagrams with an increasing number of rungs, corresponding to a certain number $n$ of exchanged gravitons \cite{Sundborg:1988tb} (\cf~Fig.~\ref{fig:Ladder-Diagram} for a relevant diagram). The saddle-point contributions evaluated at $b = b_{*}$ therefore contains sums over $n$ gravitons
\begin{align}
	A^{\text{E}}
		&\sim
								\exp\!
								\big[\.
									\irm\.c\.s^{\alpha} \theta^{\beta}
								\big]
		=
								\sum_{n} \.\frac{ \left( \irm\.c\.s^{\alpha} \theta^{\beta} \right)^{n} }{ n! }
								\; ,
								\label{eq:exponent}
\end{align}
where $c$ is a constant.

We would like to derive the value of $n$ which gives the dominant contribution. For large $s$ we expect it to be large, and using Stirling's approximation we estimate the $n$'th term as
\begin{align}
	\frac{ 1 }{ \sqrt{2 \pi\.n\.}\.( n / e )^{n} }\.
	\exp\!
	\Big[
		n\.\ln\!
		\big(
			\irm\.c\.s^{\alpha}\theta^{\beta}
		\big)
	\Big]
		&\sim
								\exp\!
								\Big[
									-
									n \ln( n )
									+
									n
									+
									n\.\ln\!
									\big(
										\irm\.c\.s^{\alpha}\theta^{\beta}
									\big)
								\Big]
								\; ,
								\label{eq:exponent-nth-term-estimate}
\end{align}
which is stationary at
\vs{-2mm}
\begin{align}
	n
		&\equiv
								n_{*}
		\sim
								\irm\.c\.s^{\alpha}\theta^{\beta}
								\; .
								\label{eq:nstar}
\end{align}
Thus we read off $\alpha$ and $\beta$ from Eq.~\eqref{eq:Born-Amplituide-large-b} and get
\begin{align}
	n_{*}
		&\sim
								\irm\.c\.s^{\frac{ D - 2 }{ 2 D - 6 }}{\theta}^{\frac{D-4}{D-3}}
								\; .
								\label{eq:nstar=N}
\end{align}
The imaginary nature of the saddle point in $n$ complicates a concrete interpretation of a dominant term in the series, but the scaling with $s$ is still significant.

In fact, we can compare this calculation to an estimate of the average number $N$ of particles of wavelength $\lambda = b_{*}$ from the ratio of the total energy $E = \sqrt{s\.}$ to the energy $\epsilon = 1 / b_{*}$ of one quantum,
\begin{align}
	N
		&\sim
								E / \epsilon
		=
								b_{*}\.\sqrt{s\.}
		\sim
								s^{\frac{ D - 2 }{ 2 D - 6 }}
		\sim
								b_{*}^{ D - 2 }
		\sim
								| n_{*} |
								\; ,
								\label{eq:N-of-s}
\end{align}
which scales as in Eq.~\eqref{eq:nstar=N}\footnote{Note that we assume that $q^{2} \sim s$, \ie~fixed angle at high energy, and only match the energy scaling.}. These estimates are precisely the same as in the quantum $N$-portrait \cite{Dvali} of black holes, \ie~we have in particular
\vs{-1mm}
\begin{align}
	b_{*}
		&\sim
								N^{\frac{ 1 }{ D - 2 }}
		\sim
								r_{\text{S}}
								\; .
								\label{eq:bstar-sim-rstar}
\end{align}
Thus, the picture of a quantum black hole as a condensate of $N$ weakly interacting gravitons is supported by and supports our interpretation of the leading high-energy scattering at small angles as an exchange of $N$ independent gravitons.

The phase in Eq.~\eqref{eq:exponent} is a constant and just ensures the right relative phase in the coherent sum over different numbers of exchanged gravitons in the eikonal amplitude (\ref{eq:Born-Amplituide-large-b},~\ref{eq:exponent}). We can study the amplitude as a sum over $n$ graviton contributions. We have already established the saddle point contribution, but the width around the saddle-point Gaussian is also useful. Differentiating the exponent in Eq.~\eqref{eq:exponent-nth-term-estimate} twice we find a width
\vs{-1mm}
\begin{align}
	\Delta n
		&\sim
								N^{- \frac{ 1 }{ 2 }}
								\; .
								\label{eq:n-width}
\end{align}
Here we have again suppressed the phase, because it is immaterial to the $n$ dependence. Comparing $\Delta n$ and $n_{*}$ from Eq.~\eqref{eq:N-of-s} we find the typical relations of a Poisson distribution of $n$. This hint motivates second look at the sum \eqref{eq:exponent}, which indeed is a Poisson distribution up to normalisation. The effective numbers of gravitons exchanged in high-energy scatterings is thus Poisson distributed with a mean growing like $N$, the numbers of gravitons in a Bose-Einstein condensate description of black holes.

Poisson distributions are characteristic of quantum mechanical coherent states, generated by classical sources, and here the situation is similar. The many-graviton state of the graviton field which is effectively exchanged between the two high-energy particles depends on the energy and the scattering angle, and with increasing energy these sources are more and more classical. The power of the energy is due to the stationary-phase effect of coherently repeated Born-term interactions weighted by the energy-dependent gravitational coupling. Thus it takes non-linearities into account although in a rudimentary way. As long as the scattering process is dominated by large distance, other non-linearities and loop effects are suppressed. This is the regime of graviton dominance and classicalization. The saddle-point estimate of the eikonal amplitude is effectively encoded by the exchange of a coherent state of the graviton field, which classically corresponds to a black hole.

\section{Summary and Discussion}
\label{sec:Summary-and-Discussion}

By reanalysing gravitational high-energy scattering we have found a confluence of pictures. A saddle point contribution may be interpreted as virtual black holes being exchanged between high-energy particles scattering at small angles; resolving these black holes in terms of perturbation theory, a Poisson distribution in the number of gravitons is found with a mean $N$ growing with energy; the gravitons are predominantly of low energy and are non-interacting at leading order. In all respects that we have checked we find support for the quantum $N$-portrait of Dvali and Gomez \cite{Dvali}. Furthermore, the scaling of $N$ with energy is precisely the scaling \cite{Sundborg:1988tb} emerging from classicalon arguments \cite{Dvali:2010jz, Dvali:2011th} requiring a balance between linear and non-linear contributions to the energy. In short, the exchanged black holes are classicalons built of Bose-Einstein condensates of gravitons.

A number of qualifying remarks are in order. First, scattering depends not only on the total energy, but also on scattering angle for elastic $2 \to 2$ amplitudes, and on more parameters in general. We see a dependence of the exchanged graviton condensate on the scattering angle, and for more complicated amplitudes we expect more complicated parameter dependence. We don't view this non-uniqueness as a failure of the picture, instead we regard the parameters of the scattering amplitude as probing different perspectives on graviton condensates. It would be extremely interesting if recent progress on amplitude technology \cite{Bern, Elvang:2013cua} would help analysing graviton condensates. Techniques for computing ladders of scalars \cite{Bastianelli:2014bfa} could perhaps also be extended to the gravitational setting. Second, fluctuations around the saddle point contributions are crucial for the quantum description of black holes. In particular, it seems clear that the states that appear naturally in amplitude calculations are coherent states rather than fixed occupation number states of the graviton field. On one hand this issue is just a choice of base, on the other hand it is probably more effective to discuss states that are easy to prepare from natural initial conditions. It could be that a better understanding of the coherent states that are excited in scattering will prove useful. Third, the connections we have made between the various pictures above have concerned the exchange of virtual states, not physical on-shell states. It would of course be extremely satisfying if the spectrum of on-shell states were understood.

\acknowledgements
This work was supported by the Swedish Research Council (VR) through the Oskar Klein Centre.



\begin{thebibliography}{10}

\bibitem{Dvali}
	G.~Dvali and C.~Gomez,
	{\it Black Hole's Quantum N-Portrait},
	Fortsch.~Phys.~{\bf 61}, 742 (2013);
	G.~Dvali and C.~Gomez,
	{\it Black Holes as Critical Point of Quantum Phase Transition},
	arXiv:1207.4059 [hep-th];
	G.~Dvali and C.~Gomez,
	{\it Black Hole Macro-Quantumness},
	arXiv:1212.0765 [hep-th];
	G.~Dvali and C.~Gomez,
	{\it Black Hole's Information Group},
	arXiv:1307.7630 [hep-th];
	G.~Dvali and C.~Gomez,
	{\it Quantum Compositeness of Gravity: Black Holes, AdS and Inflation},
	arXiv:1312.4795 [hep-th];
	G.~Dvali and C.~Gomez,
	{\it Black Hole's {$1 / N$} Hair},
	Phys.~Lett.~B {\bf 719}, 419 (2013);
	G.~Dvali and C.~Gomez,
	{\it BICEP2 in Corpuscular Description of Inflation},
	arXiv:1403.6850 [astro-ph.CO].

\bibitem{N-Portrait-related}
	F.~Berkhahn, S.~M{\"u}ller, F.~Niedermann and R.~Schneider,
	{\it Microscopic Picture of Non-Relativistic Classicalons},
	arXiv:1302.6581 [hep-th];
	D.~Flassig, A.~Pritzel and N.~Wintergerst,
	{\it Black Holes and Quantumness on Macroscopic Scales},
	Phys.~Rev.~D {\bf 87} 084007 (2013);
	G.~Dvali, D.~Flassig, C.~Gomez, A.~Pritzel and N.~Wintergerst,
	{\it Scrambling in the Black Hole Portrait},
	arXiv:1307.3458 [hep-th].

\bibitem{Brustein}
	R.~Brustein,
	{\it Origin of the blackhole information paradox},
	Fortsch.~Phys.~{\bf 62}, 255 (2014);
	R.~Brustein and A.~J.~M.~Medved,
	{\it Horizons of semiclassical black holes are cold},
	arXiv:1312.0880 [hep-th];
	R.~Brustein and A.~J.~M.~Medved,
	{\it Phases of information release during black hole evaporation},
	JHEP {\bf 1402}, 116 (2014).

\bibitem{Casadio}
	R.~Casadio and A.~Orlandi,
	{\it Quantum Harmonic Black Holes},
	JHEP {\bf 1308}, 025 (2013).
	R.~Casadio, A.~Giugno, O.~Micu and A.~Orlandi,
	{\it Black holes as self-sustained quantum states, and Hawking radiation},
	arXiv:1405.4192 [hep-th].

\bibitem{Sundborg:1988tb}
	B.~Sundborg,
	{\it High-energy Asymptotics: The One Loop String Amplitude and Resummation},
	Nucl.~Phys.~B {\bf 306}, 545 (1988).


\bibitem{Amati}
	D.~Amati, M.~Ciafaloni, G.~Veneziano,
	{\it Superstring collisions at planckian energies},
	Phys.~Lett.~B {\bf197}, 81 (1987).
	
\bibitem{Muzinich:1987in}
	I.~J.~Muzinich and M.~Soldate,
	{\it High-Energy Unitarity of Gravitation and Strings},
	Phys.~Rev.~D {\bf 37}, 359 (1988).

\bibitem{Giddings}
	S.~B.~Giddings, D.~J.~Gross and A.~Maharana,
	{\it Gravitational effects in ultrahigh-energy string scattering},
	Phys.~Rev.~D {\bf 77}, 046001 (2008);
	S.~B.~Giddings and M.~Srednicki,
	{\it High-energy gravitational scattering and black hole resonances},
	Phys.~Rev.~D {\bf 77}, 085025 (2008);
	S.~B.~Giddings, M.~Schmidt-Sommerfeld and J.~R.~Andersen,
	{\it High energy scattering in gravity and supergravity},
	Phys.~Rev.~D {\bf 82}, 104022 (2010).

\bibitem{Giddings:2011xs}
	S.~B.~Giddings,
	{\it The gravitational S-matrix: Erice lectures},
	arXiv:1105.2036 [hep-th].

\bibitem{D'Appollonio:2010ae}
	G.~D'Appollonio, P.~Di Vecchia, R.~Russo and G.~Veneziano,
	{\it High-energy string-brane scattering: Leading eikonal and beyond,}
	JHEP {\bf 1011}, 100 (2010);
	G.~D'Appollonio, P.~Di Vecchia, R.~Russo and G.~Veneziano,
	{\it The leading eikonal operator in string-brane scattering at high energy},
	arXiv:1310.4478 [hep-th].

\bibitem{t'Hoof1987}
	G.~'t Hooft,
	{\it Graviton Domination in Ultra-High-Energy-Scattering},
	Phys.~Lett.~B {\bf198}, 61 (1987).

\bibitem{Dvali:2010jz}
	G.~Dvali, C.~Gomez and A.~Kehagias,
	{\it Classicalization of Gravitons and Goldstones},
	JHEP {\bf 1111}, 070 (2011);
	G.~Dvali and D.~Pirtskhalava,
	{\it Dynamics of Unitarization by Classicalization},
	Phys.~Lett.~B {\bf 699}, 78 (2011);
	G.~Dvali,
	{\it Classicalize or not to Classicalize?},
	arXiv:1101.2661 [hep-th].

\bibitem{Dvali:2011th}
	G.~Dvali, G.~F.~Giudice, C.~Gomez and A.~Kehagias,
	{\it UV-Completion by Classicalization},
	JHEP {\bf 1108}, 108 (2011).

\bibitem{Dvali:2010bf}
	G.~Dvali and C.~Gomez,
	{\it Self-Completeness of Einstein Gravity},
	arXiv:1005.3497 [hep-th].

\bibitem{Kuhnel:2014zja}
	F.~Kuhnel and B.~Sundborg,
	{\it ``Decay of Graviton Condensates and their Generalizations in Arbitrary Dimensions},
	arXiv:1405.2083 [hep-th].

\bibitem{Collins}
	P.D.B.~Collins,
	{\it An introduction to Regge theory and high energy physics},
	Cambridge University Press,
	Cambridge (1977), and references therein.

\bibitem{Bern}
	Z.~Bern,
	{\it Perturbative quantum gravity and its relation to gauge theory},
	Living Rev.~Rel.~{\bf 5}, 5 (2002);
	Z.~Bern, J.~J.~M.~Carrasco and H.~Johansson,
	{\it New Relations for Gauge-Theory Amplitudes},
	Phys.~Rev.~D {\bf 78}, 085011 (2008);
	Z.~Bern, T.~Dennen, Y.~-t.~Huang and M.~Kiermaier,
	{\it Gravity as the Square of Gauge Theory},
	Phys.~Rev.~D {\bf 82}, 065003 (2010);
	Z.~Bern, J.~J.~M.~Carrasco and H.~Johansson,
	{\it Perturbative Quantum Gravity as a Double Copy of Gauge Theory},
	Phys.~Rev.~Lett.~{\bf 105}, 061602 (2010).

\bibitem{Elvang:2013cua}
	H.~Elvang and Y.~-t.~Huang,
	{\it Scattering Amplitudes},
	arXiv:1308.1697 [hep-th].

\bibitem{Bastianelli:2014bfa}
	F.~Bastianelli, A.~Huet, C.~Schubert, R.~Thakur and A.~Weber,
	{\it Integral representations combining ladders and crossed-ladders},
	arXiv:1405.7770 [hep-ph].	

\end{thebibliography}
\end{document}